\documentclass[amsmath,amssymb]{iopart}

\usepackage{graphicx}
\usepackage{dcolumn}
\usepackage{bm}
\usepackage{amsfonts}
\begin{document}


\title{Spectral inversion of an indefinite Sturm-Liouville problem due to Richardson}

\author{Paul E. Shanley}
\address{Department of Physics, University of Notre Dame, Notre Dame, IN 46556, USA.}
\ead{pshanley@nd.edu}


\begin{abstract}
We study an indefinite Sturm-Liouville problem due to Richardson whose complicated eigenvalue dependence on a parameter has been a puzzle for decades.  In atomic physics a process exists that inverts the usual Schr\"{o}dinger situation of an energy eigenvalue depending on a coupling parameter into the so-called Sturmian problem where the coupling parameter becomes the eigenvalue which then depends on the energy.  We observe that the Richardson equation is of the Sturmian type.  This means that the Richardson and its related Schr\"{o}dinger eigenvalue functions are inverses of each other and that the Richardson spectrum is therefore no longer a puzzle.
\end{abstract}

\maketitle

\section{\label{sec:Intro}Introduction}

In the early years of the twentieth century an eigenvalue problem based on an indefinite second order differential equation of the Sturm-Liouville(SL) type was introduced by the mathematician R. G. D. Richardson \cite{Richardson}.  The history of this early period has been reviewed by Mingarelli \cite{Mingarelli} with later developments summarized by Binding and Volkmer \cite{Binding}.  The spectrum of the Richardson problem depends on a real parameter in a very complicated way that has never been fully understood and has even been called amazing \cite{Atkinson}.  Our goal here is to provide a non-rigorous explanation of this spectrum.

Although the Richardson problem predates quantum mechanics, we make a quantum connection by asserting that, in the language of atomic physics, it is a Sturmian problem \cite{Avery}.  In such a view the Richardson eigenvalues are coupling parameters of a potential and the parameter that causes the complication is an energy.  In the atomic physics of the Coulomb potential one is motivated to go from Schr\"{o}dinger to Sturmian for computational advantage.  Here we reverse the direction and go from Richardson as a Sturmian problem to Richardson as a Schr\"{o}dinger problem where now more naturally the parameter is the coupling and the energy is the eigenvalue.  It is then seen that the Richardson and Schr\"{o}dinger eigenvalue functions are inverses of each other and thus both contain the same information.  We are then led to understand Richardson in terms of Schr\"{o}dinger.

In the Richardson problem the independent variable is the energy $E$ and we call the dependent ones the eigencouplings $\lambda_n(E)$.  In this case for $E$ real and fixed we have one energy and many potentials while in the Schr\"{o}dinger problem with $\lambda$ fixed we have one potential and many energies $E_n(\lambda)$.  We use the term spectral inversion to mean going from one description to the other.  The complications of the Richardson spectrum \cite{Atkinson} involve short segments in $E$ separated by square root branch points in which $\lambda_n(E)$ is alternatively real, imaginary or complex, and these possibilities recur over and over as $E$ varies through real values.  In the related Schr\"{o}dinger problem, $E_n(\lambda)$ for real $\lambda$ is much simpler and this difference in complication for real values of the parameters must stem from a difference in the two SL equations that generate the spectra.  In SL terminology \cite{Zettl} the Schr\"{o}dinger equation is intrinsically right-definite and if certain technical requirements are satisfied, the $E_n(\lambda)$ are real for real coupling. In the Richardson case the eigencoupling multiplies the spacial part of the potential and if this function changes sign as it does in the Richardson problem, the SL equation can be both left- and right-indefinite and this leads to a spectrum that is both non-real and singular as $E$ varies through real values.  In this paper we argue that one of the two descriptions may not be simpler than the other because inverse functions contain the same information.  This equality implies that the Schr\"{o}dinger version must be suitably complicated in the complex domain of $\lambda$ where the $E_n(\lambda)$ are less tightly constrained. In this way a balance of complication can be maintained. It must be so that the Schr\"{o}dinger complication is more subtle and is not as evident in its SL equation.

To examine these questions we need to study the eigenquantities in the complex domains of their independent variables, or more generally on the Riemman surfaces on which they take their values.   The analytic properties are important and we need to focus on their square root branch points and critical points since one type of point becomes the other under spectral inversion.  It will turn out that critical points are the sources of the Schr\"{o}dinger complication while square root branch points are responsible for the complication in the Richardson problem.  Thus spectral inversion relates one source of complication to the other.  From the mathematical point of view it also relates an indefinite SL problem to a definite one.
\noindent

In section \ref{sec:Level} we relate the critical and square root branch point of the Richardson and Schr\"{o}dinger problems in terms of single and double zeros of a certain spectral function.  It is also pointed out that Schr\"{o}dinger critical points for real $\lambda$ lead to curves in the upper and lower planes of $\lambda$ on which the eigenvalues are real.  Section \ref{sec:Rich} discusses the Richardson and Schr\"{o}dinger differential equations. In section \ref{sec:Diff} we treat the problems more generally and develop equations for the generalized eigenfunctions that are present at branch points. The Schr\"{o}dinger spectrum is studied in detail in section \ref{sec:Schro} where branch points and critical points are located and the Riemann surface structure is discussed.  Section \ref{sec:Richspec} treats the analogous properties of the Richardson spectrum and the structure of its Riemann surface. A qualitative demonstration that the inversion of the Richardson $\lambda_n(E)$ leads to the Schr\"{o}dinger $E_n(\lambda)$ is presented in section \ref{sec:Spect}.  Some questions concerning oscillation counts and defective multiplicities are treated in section \ref{sec:Oscil}.  Things are summarized in section \ref{sec:Sum}.

\section{\label{sec:Level}Level crossing and its analog}

Our goal in the section is to show the relation between singular points and critical points in the Schr\"{o}dinger and Sturmian descriptions.  In the mathematical literature there is some discussion of these questions at an abstract \cite{Binding2} as well as at a more concrete \cite{Binding} level.  

We start by considering a function $\Delta(\lambda,E)$ of the coupling parameter $\lambda$ and energy $E$ whose vanishing indicates a discrete level of the system.  For example, this function could be the secular determinant formed in an attempted diagonalization of the Hamiltonian in some basis.  The vanishing allows us to define an energy $E(\lambda)$ if $\lambda$ is independent or the eigencoupling $\lambda(E)$ if $E$ is independent.
\noindent
To be specific in the Schr\"{o}dinger case suppose that values $\lambda_0$ and $E_0$ cause $\Delta$ to vanish. Expanding near the zero we have

\vspace{0.3cm}
\begin{equation}
\Delta(\lambda,E)=\Delta_\lambda(\lambda-\lambda_0)+\Delta_E(E-E_0)+\cdots,
\label{eq:Delta}
\end{equation}\vspace{0.3cm}
with the subscripts indicating partial derivatives. Choosing $\lambda$ so that $\Delta(\lambda,E)$ vanishes allows a local solution for $E(\lambda)$

\vspace{0.3cm}
\begin{equation}
E(\lambda)=E(\lambda_0)-\frac{\Delta_\lambda}{\Delta_E}(\lambda-\lambda_0)+\cdots=E(\lambda_0)+\frac{dE}{d\lambda}
\bigg|_{\lambda_0}(\lambda-\lambda_0)+\cdots .
\label{eq:dellam}
\end{equation}\vspace{0.3cm}
Level crossing at $\lambda_0$ means that $\Delta$ has a double root so that $\Delta_E$ also vanishes which makes $E(\lambda)$ singular.  This argument does not disclose the nature of the singularity but generically they are square roots which we will assume to be the case since we have no numerical evidence to the contrary.

If the above steps are repeated for the Sturmian problem and if $\Delta(\lambda_1,E_1)=0$, we obtain

\vspace{0.3cm}
\begin{equation}
\lambda(E)=\lambda(E_1)-\frac{\Delta_E}{\Delta_\lambda}(E-E_1)+\cdots=\lambda(E_1)+\frac{d\lambda}{dE}
\bigg|_{E_1}(E-E_1)+\cdots.
\label{eq:delE}
\end{equation}\vspace{0.3cm}
Coupling crossing occurs if $\Delta_\lambda=0$ which makes $\lambda(E)$ singular.  The possibility of coincident zeros of $I_E$ and $I_\lambda$ has been discussed and eliminated at least in one case \cite{Simon}.

We now return to the Schr\"{o}dinger problem and ask what property of $E(\lambda)$ at a regular point of $\lambda$ implies $\lambda(E)$ is singular at the image point. Returning to (\ref{eq:dellam}) we invoke $\Delta_E\neq0$ with $\Delta_\lambda=0$ which gives $E'(\lambda)=0$. Thus a critical point of $E(\lambda)$ implies that $\lambda(E)$ is singular.  Of course a critical point position $\lambda_0$ and critical value $E_0(\lambda_0)$ in the Schr\"{o}dinger description also gives the branch point position $E_0$ and singular value $\lambda_0(E_0)$ in the other description.  In a similar way in (\ref{eq:delE}), a critical point of $\lambda(E)$ signals level crossing in $E(\lambda)$.  We assume the absence of triple or higher crossing in a problem with one coupling parameter present which implies that the critical points are non-degenerate so that second derivatives are non-vanishing at the points.  In what follows we will see that the eigenquanties in the two problems take their values on Riemann surfaces so that when we go from one to the other a (regular non-critical, regular critical, singular)point will become a (regular non-critical, singular, regular critical) point on the other.  We will refer to the Schr\"{o}dinger Riemann surface as $\mathbb{R}_S$ and to the coupling one of $\lambda_n(E)$ as $\mathbb{R}_C$.

We have stated that critical points are the sources of Schr\"{o}dinger complication and this arises from the leading quadratic behavior in their vicinity.  This induces a more rapid angular dependence and causes contours of $\textrm{Re}~ E_n(\lambda)=\textrm{constant}$ and of $\textrm{Im}~ E_n(\lambda)=\textrm{constant}$ to intersect in a double-x fashion.  If $E_n(\lambda)$ is real and has a critical point for $\lambda$ real, it acts as a source of secondary curves into the upper and lower planes on which $E_n(\lambda)$ is also real.  The real contour and the real axis are locally orthogonal but the curve then bends away depending on local details.  Problems exhibiting PT-symmetry \cite{Bender} usually have potentials with an even-odd spacial structure that leads to $E_n(-\lambda)=E_n(\lambda)$.  This makes $\lambda=0$ a critical point but also forces all higher odd derivatives to vanish causing curves into the upper and lower plane to coincide with the imaginary axis over an extended interval.  It is this special type of critical point that gives rise to the properties of PT-symmetry.  If any odd derivative higher than the first is non-vanishing, then the curves will not coincide with the axis and the eigenvalues will be non-real on it.  If one is interested in real $E_n(\lambda)$ in the complex domain of $\lambda$ where the Schr\"{o}dinger operator is not self-adjoint, critical points of $E_n(\lambda)$ for real $\lambda$ are of special importance.  Under spectral inversion these points will become square root singularities for $E$ real and thus contribute to making $\lambda_n(E)$ complicated.

\section{\label{sec:Rich}Richardson and Schr\"{o}dinger problems}

The Richardson problem is classed as a right-indefinite SL equation and may be written

\vspace{0.3cm}
\begin{eqnarray}
-\phi_n^{''}(E,x)-E\phi_n(E,x)=\lambda_n(E)~\textrm{sgn}(x)\phi_n(E,x),\label{eq:Richa}\\
\begin{array}{l}\phi_n(E,\pm1)=0\\-1\leq x\leq1\\\end{array}\begin{array}{c}~~~~~~~~~~~~~~~~\end{array}
\begin{array}{r}\textrm{sgn}(x)=\bigg\{\begin{array}{c}-1,x<0\\+1,x>0\end{array}\end{array}.\nonumber \\ \nonumber
\end{eqnarray}
$E$ is a parameter that is usually real and the function $\textrm{sgn}(x)$ changes sign once in the interval and is termed an indefinite weight.  Its presence leads to the complication of the Richardson spectrum; if it is replaced by 1, the problem is elementary.  The eigencouplings $\lambda_n(E)$ are the quantities of primary interest.

When we switch to the Schr\"{o}dinger description, the SL convention is to put the eigenvalue term on the right so that $E\rightarrow E_n(\lambda), \lambda_n(E)\rightarrow \lambda$, and we obtain

\vspace{0.3cm}
\begin{eqnarray}
-\psi_n^{''}(\lambda,x)-\lambda~\textrm{sgn}(x)\psi_n(\lambda,x)=E_n(\lambda)\psi_n(\lambda,x),\label{eq:Richb}\\
\begin{array}{l}-1\leq x\leq1\\\end{array}\begin{array}{c}~~~~~~~~~~~~~~~~\end{array}
\begin{array}{r}\psi_n(\lambda,\pm1)=0\end{array}.\nonumber\\ \nonumber
\end{eqnarray}
The indefinite weight has become incorporated into the potential $V(x)=-\lambda~\textrm{sgn}(x)$ and may seem to be of less significance; no matter how many times $V(x)$ changes sign, (\ref{eq:Richb}) is right-definite and is much better behaved than (\ref{eq:Richa}).  This potential for purely imaginary coupling has been independently suggested as a problem exhibiting PT-symmetry by Znojil \cite{Znojil}, followed by joint work with L\'{e}vai \cite{Znojil2} and we can compare our results to theirs in Section \ref{sec:Schro}.

We now make a number of remarks concerning spectral inversion. If we go from Richardson to Schr\"{o}dinger we have $\lambda_n(E)\rightarrow E_n(\lambda)$ so that place~$\rightarrow$~value and value~$\rightarrow$~place. The Richardson $\lambda_n(E)$ represent complicated values and real places which when inverted become real values and complicated places.  This Schr\"{o}dinger combination has as its source the critical points for real $\lambda$ that produce curves in the upper and lower plane on which $E_n(\lambda)$ is real.  It follows that we view the Richardson spectrum, we are viewing the complex geography of these Schr\"{o}dinger curves.

Suppose we have determined that particular values of $\lambda_0$ and $E_0$ are image points on the two Riemann surfaces. If we insert these numbers into equations (\ref{eq:Richa}) and (\ref{eq:Richb}), the left-right details are irrelevant and the equations coincide so that the eigenfunctions can be taken to be identical.  In the real domains of $\lambda$ and $E$ if we let the oscillation count denote the number of internal nodes of either eigenfunction, then that count as determined at image points is invariant under spectral inversion, as is the eigenfunction.

\section{\label{sec:Diff}Differential equation}

We need to examine the differential equations of the two problems in more detail so that the role of the critical and branch points can be made clear.  We also study the generalized eigenfunctions present at branch points where either spectrum is defective. In the Schr\"{o}dinger case Simon \cite{Simon} has shown in a study of the anharmonic oscillator that the inhomogeneous equation for the generalized eigenfunction can be obtained by taking the derivative of the Schr\"{o}dinger equation with respect to $E$.  In both  Schr\"{o}dinger and Richardson cases we will use procedures analogous to this.

A more general potential $V(x)=\lambda v(x)+w(x)$ is chosen where the two functions are real and $\lambda$ may be complex.  Initially, we leave open whether $\lambda$ or $E$ is independent.  In abbreviated notation the Schr\"{o}dinger equation is

\vspace{0.3cm}
\begin{eqnarray}
-\psi^{''}(x)+\lambda v(x)\psi(x)+w(x)\psi(x)-E\psi(x)=0,\label{eq:Diff1}\\
\begin{array}{l}\psi(a)=\psi(b)=0,\\\end{array}\begin{array}{c}~~~~~~~~~~\end{array}
\begin{array}{r}-\infty< a\leq x\leq b<\infty\end{array}.\nonumber\\ \nonumber
\end{eqnarray}
In addition to this it is useful to have the following result.  Suppose a second function $\xi(x)$ obeys an equation that is identical to (\ref{eq:Diff1}) on the left but has an inhomogeneous term $F(x)$

\begin{eqnarray}
-\xi^{''}(x)+\lambda v(x)\xi(x)+w(x)\xi(x)-E\xi(x)=F(x),\label{eq:Diff2}\\
\xi(a)=\xi(b)=0.\nonumber\\ \nonumber
\end{eqnarray}
We want a solution to both equations for the same values of $\lambda$ and $E$.  If we multiply (\ref{eq:Diff1}) by $\xi(x)$ and (\ref{eq:Diff2}) by $\psi(x)$, subtract and integrate over $x$ from $a$ to $b$ we obtain

\vspace{0.3cm}
\begin{equation}
\langle F|\psi\rangle_R=\int_a^bdxF(x)\psi(x)=0.
\label{eq:Diff3}
\end{equation}\vspace{0.3cm}
In other words for a pair of solutions to exist, the inhomogeneous term must be orthogonal to the homogeneous solution with respect to a real ($R$) inner product.

We now go to a point $E$ on $\mathbb{R}_C$ and wish to study the derivative $\partial_E\psi(x)=\psi_E(x)$.  Letting $\mathfrak{D}=-\partial_x^2+\lambda v(x)+w(x)-E$ and acting on (\ref{eq:Diff1}) with $\partial_E$ we get the pair

\begin{eqnarray}
&\mathfrak{D}\psi(x)&= 0,\\
&\mathfrak{D}\psi_E(x)&= -(\partial\lambda/\partial E)v(x)\psi(x)+\psi(x).\nonumber\\ \nonumber
\label{eq:Diff4}
\end{eqnarray}\vspace{0.3cm}
At a critical point the derivative on the right vanishes giving the simpler pair

\begin{eqnarray}
\mathfrak{D}\psi(x)&=& 0,\label{eq:Diff5}\\
\mathfrak{D}\psi_E(x)&=& \psi(x).\nonumber\\ \nonumber
\end{eqnarray}
Solution of these equations would produce four things: the critical point position $E$  on $\mathbb{R}_C$, the critical value $\lambda(E)$, as well as $\psi(x)$ and $\psi_E(x)$.  The image branch point on $\Re_S$ has the same values of $\lambda$ and $E$, $\partial\lambda/\partial E$ vanishes there as well, and (\ref{eq:Diff5}) has the standard form \cite{Simon} for a eigenfunction-generalized eigenfunction pair at a branch point in a Schr\"{o}dinger problem.  We then associate this $\psi_E(x)$ with the generalized eigenfunction on $\mathbb{R}_S$. Putting in a full notation and renaming $\psi_E\rightarrow\tilde{\psi}$, the pair (\ref{eq:Diff5}) becomes

\vspace{0.3cm}
\parbox{9cm}
{$\begin{array}{l}\\
-\psi_n^{''}(\lambda,x)+\lambda v(x)\psi_n(\lambda,x)+w(x)\psi_n(\lambda,x)= E_n(\lambda)\psi_n(\lambda,x),\\ 
\\
-\tilde{\psi_n}^{''}(\lambda,x)+\lambda v(x)\tilde{\psi}_n(\lambda,x)+w(x)\tilde{\psi}_n(\lambda,x)-E_n(\lambda)\tilde{\psi}_n(\lambda,x)= \psi_n(\lambda,x).\\
\\
\end{array}$}
\hfill\parbox{3cm}
{\begin{eqnarray}\label{eq:Diff6}\end{eqnarray}} \vspace{0.3cm}
The constraint of (\ref{eq:Diff3}) gives

\vspace{0.3cm}
\begin{equation}
\langle \psi_n(\lambda)|\psi_n(\lambda)\rangle_R=\int_a^bdx\psi_n(\lambda,x)\psi_n(\lambda,x)=0.
\label{eq:Diff7}
\end{equation}\vspace{0.3cm}

\noindent
We can carry out an analogous procedure starting at a critical point $\lambda$ on $\mathbb{R}_S$.  We write $\psi_\lambda(x)=\partial_\lambda\psi(x)$ and apply the derivative to (\ref{eq:Diff1}) which gives

\begin{eqnarray}
\mathfrak{D}\psi(x)&=& 0,\label{eq:Diff8}\\
\mathfrak{D}\psi_\lambda(x)&=& (\partial E/\partial\lambda)\psi(x)-v(x)\psi(x).\nonumber\\ \nonumber
\end{eqnarray}
Whether at the critical point on $\mathbb{R}_S$  or the image branch point on $\mathbb{R}_C$, $\partial E/\partial\lambda=0$ and the simpler pair is

\begin{eqnarray}
\mathfrak{D}\psi(x)&=& 0,\\
\mathfrak{D}\psi_\lambda(x)&=& -v(x)\psi(x).\nonumber\\ \nonumber
\label{eq:Diff9}
\end{eqnarray}
As before, $\psi_\lambda(x)$ has the interpretation as the generalized eigenfunction on $\mathbb{R}_C$.  We rename $\psi\rightarrow\phi$ and $\psi_\lambda\rightarrow\tilde{\phi}$ to give the pair of equations

\parbox{9cm}
{$\begin{array}{l}\\ 
-\phi_n^{''}(E,x)+ w(x)\phi_n(E,x)-E\phi_n(E,x)= -\lambda_n(E)v(x)\phi_n(E,x),\\
\\
-\tilde{\phi_n}^{''}(E,x)+w(x)\tilde{\phi}_n(E,x)-E\tilde{\phi}_n(E,x)+\lambda_n(E) v(x)\tilde{\phi}_n(E,x)\\
~~~~~~~~~~~~~~~~~~~~~~~~~~~~~~~~~~~~~~~~~~~~~~~~~~~~~~~~~~~~= -v(x)\phi_n(E,x).\\
\end{array}$}\hfill\parbox{3.5cm}{\begin{eqnarray}\label{eq:Diff10}\end{eqnarray}}
Imposing the constraint of (\ref{eq:Diff3}) gives

\vspace{0.3cm}
\begin{equation}
\langle \phi_n(E)|v|\phi_n(E)\rangle_R=\int_a^bdx\phi_n(E,x)v(x)\phi_n(E,x)=0.
\label{eq:Diff11}
\end{equation}\vspace{0.3cm}

The quantity $\psi_\lambda(x)$ at a point $\lambda$ is the first order correction to $\psi(x)$ in Rayleigh-Schr\"{o}dinger perturbation theory.  If one calculates this at a critical point on $\mathbb{R}_S$ by the usual procedures, one has wittingly or unwittingly also calculated the generalized eigenfunction at the image branch point on $\mathbb{R}_C$.

For non-critical and non-singular points in either problem, the second equation in each pair is ignored and we deal only with the first of (\ref{eq:Diff6}) or the first of (\ref{eq:Diff10}). The Schr\"{o}dinger version is intrinsically right-definite while the Sturmian one is right-indefinite if $v(x)$ changes sign.

Although we cannot write expressions for $E_n(\lambda)$ or $\lambda_n(E)$, we can obtain useful ones for their derivatives by the following procedure.  We return to the pair (\ref{eq:Diff8}) valid at a point $\lambda$ on $\mathbb{R}_S$.  The inhomogeneous term must conform to (\ref{eq:Diff3}) which permits a solution for the derivative

\vspace{0.3cm}
\begin{equation}
\frac{\partial E_n (\lambda)}{\partial\lambda}=\frac{\langle\psi_n(\lambda)|v|\psi_n(\lambda)\rangle_R}{\langle\psi_n(\lambda)|\psi_n(\lambda)\rangle_R}.
\label{eq:Diff12}
\end{equation}\vspace{0.3cm}
Going to the other pair (\ref{eq:Diff4}) and performing a similar step gives

\vspace{0.3cm}
\begin{equation}
\frac{\partial\lambda_n (E)}{\partial E}=\frac{\langle\phi_n(E)|\phi_n(E)\rangle_R}{\langle\phi_n(E)|v|\phi_n(E)\rangle_R}.
\label{eq:Diff13}
\end{equation}\vspace{0.3cm}
If $\lambda$ and $E$ are image points, we can take $|\psi_n(\lambda)\rangle=|\phi_n(E)\rangle$ and the two expressions say simply that $E'(\lambda)=1/\lambda'(E)$.  The integrals implicit in these expressions may not themselves be singular since the integrands are entire and the intervals are finite.  Singularities must come from vanishing denominators which arise from the orthogonality conditions in (\ref{eq:Diff7}) and (\ref{eq:Diff11}). In a similar way critical points arise from vanishing numerators. Although we do not use these derivative expressions in practice, they contain the essence of the relationship between the Schr\"{o}dinger and Richardson problems.

If $\lambda$ is real, the first of (\ref{eq:Diff6}) is right-definite and $E_n(\lambda)$ and $|\psi_n(\lambda)\rangle$ are real also so that the denominator in (\ref{eq:Diff12}) will not vanish and $E_n(\lambda)$ will be non-singular as expected.  According to (\ref{eq:Diff7}) the denominator will vanish at branch points which must then occur only off the real axis. If $\lambda$ is real and $v(x)$ is of one sign, the numerator cannot vanish which excludes critical points and leaves $E_n(\lambda)$ both non-singular and non-oscillatory. If $\lambda$ is real and $v(x)$ changes sign, the numerator may vanish leading to curves in the upper and lower plane on which $E_n(\lambda)$ is real.  All such points on $\mathbb{R}_S$, whether $\lambda$ be real, imaginary or complex, must map to the real axis of $\mathbb{R}_C$ on some sheet and thereby generate $\lambda_n(E)$ that match the three possibilities of $\lambda$ on $\mathbb{R}_S$.  This is how the Schr\"{o}dinger complication leads to the Richardson one.

If we attempt a similar analysis of (\ref{eq:Diff13}) for real $E$, we can draw no general conclusion because the first equation of (\ref{eq:Diff10}) may be sufficiently indefinite to give complex results. If $v(x)$ is one sign, right-definiteness should make everything real with numerator and denominator non-vanishing in (\ref{eq:Diff13}).  This gives again a non-singular and non-oscillatory function and it is natural that two such featureless functions can be inverses of each other.  Suppose for $E$ real that $v(x)$ changes sign so that the denominator vanishes giving a branch point on $\mathbb{R}_C$.  The vanishing denominator here at the point $E$ is the same as the vanishing numerator in (\ref{eq:Diff12}) at the critical point $\lambda$.  This is just another way of saying that the eigenquantities are the inversions of each other.  We conclude that sign changes in $v(x)$ can induce both Schr\"{o}dinger and Richardson complications.

There is another tool that is useful in characterizing the spectrum in each problem. If $v(x)$ is of one sign, then each eigenquantity is a Herglotz fucntion \cite{Simon} of its independent variable.  To study this we start with the first equation of (\ref{eq:Diff6}) and act on it with $\textrm{Im}\int_a^bdx\psi_n^\ast(\lambda,x)$.  The integral of the derivative term and of the $w(x)$ term are real and do not contribute, leading to

\vspace{0.3cm}
\begin{equation}
\textrm{Im} E_n(\lambda)=\textrm{Im}\lambda\frac{\langle\psi_n(\lambda)|v|\psi_n(\lambda)\rangle_H}{\langle\psi_n(\lambda)|\psi_n(\lambda)\rangle_H},
\label{eq:Diff14}
\end{equation}\vspace{0.3cm}
where the $H$ denotes Hilbert space inner product.  If $v(x)$ is positive in $(a,b)$, the ratio is real and positive for all $n$ and all complex $\lambda$. This makes $E_n(\lambda)$ a Herglotz function and it can be real only on its real axis.  Critical points would produce curves that violate this and are not allowed. The Herglotz property and PT-symmetry are therefore incompatible. If $v(x)$ was always negative, we could come to the same conclusions and refer to the function as anti-Herglotz.

To study $\lambda_n(E)$ we start from the first equation of (\ref{eq:Diff10}) and take similar steps to get

\vspace{0.3cm}
\begin{equation}
\textrm{Im}\lambda_n(E)=\textrm{Im} E\frac{\langle\phi_n(E)|\phi_n(E)\rangle_H}{\langle\phi_n(E)|v|\phi_n(E)\rangle_H}.
\label{eq:Diff15}
\end{equation}\vspace{0.3cm}
If $v(x)$ is always positive then $\lambda_n(E)$ is also Herglotz and so this property is invariant under spectral inversion.  In other words the three possibilities of upper plane, real axis, and lower plane map, respectively, into the same possibilities.

The right-definite arguments that we have used lead to real eigenquantites when their parameter is real.  If either of the problems had real curves off the real axis of their parameter, the complex places and real values would invert to real places and complex values.  Since this is impossible, both arguments rule out such curves. This argument can also be applied to the Richardson problem to exclude the existence of curves off the real axis of $E$ on which $\textrm{Im}\lambda_n(E)=0$.  The spectral inversion of this would be unacceptable to the Schr\"{o}dinger problem.

Suppose $v(x)$ does change sign so that critical points generate real eigenvalues for non-real $\lambda$.  $E_n(\lambda)$ is then not Herglotz but (\ref{eq:Diff14}) is still valid and it follows that $\langle\psi_n(\lambda)|v|\psi_n(\lambda)\rangle_H=0$ at every point on such curves.  The algebraic sign of this quantity is called the signature \cite{Binding3} and can be used to obtain oscillation results for indefinite problems.

An example of a problem that is both right- and left-definite in its Schr\"{o}dinger and Sturmian versions is a quartic anharmonic oscillator \cite{Shanley} with $V(x)=\lambda x^2+x^4$.  In this problem on $x\in[-\infty,\infty]$, the additional function $w(x)=x^4$ serves to stabilize the spectrum at $\lambda=0$. Each eigenquantity is a Herglotz function of its parameter and is therefore non-oscillatory and non-singular in the real domains.

\section{\label{sec:Schro}Schr\"{o}dinger spectrum}

In this section we focus on the eigenvalues of the Schr\"{o}dinger equation.  Critical and branch points are located and their properties studied. The parameter $E$ that is real in Richardson becomes the real $E_n(\lambda)$ in Schr\"{o}dinger.  We must then search in complex $\lambda$ for real $E_n(\lambda)$.  The Riemann surface of $E_n(\lambda)$ is also described.

Given the equation $\textrm{Im}E_n(\lambda)=0$, we ask for what complex $\lambda$ is this true?  Since $\textrm{Im}E_n(\lambda)$ is itself real, we make the replacements $\textrm{Im}E_n\rightarrow F$, $\textrm{Re}\lambda\rightarrow x$, $\textrm{Im}\lambda\rightarrow y$, giving the real equation $F(x,y)=0$.  The implicit function theorem implies that the solution is a curve and in the complex case we call such a curve a real locus and allow it to consist of the following possibilities: the real axis $A$, the imaginary axis $B$, or an actual curve $C$ not on either axis.  This is not to say that all problems have the three types but only that these are the possibilities.  A self-adjoint problem with real coupling will have $A$'s while a problem exhibiting PT-symmetry may have $B$'s.  We know of no discussion of possibility $C$ but this type must be present in the Schr\"{o}dinger problem if the Richardson spectrum is to be understood.  Since it will be argued that the Riemann surface of $E_n(\lambda)$ has an infinite number of sheets, we should expect and infinite number of each of the three types.

It is natural that an $A$ and a $B$ will intersect at $\lambda=0$ and our discussion in section \ref{sec:Level} implies that all odd derivatives of $E_n(\lambda)$ will vanish there if $E_n(-\lambda)=E_n(\lambda)$.  If some odd derivative other than the first is non-vanishing, then a $C$ would arise at $\lambda=0$ and not a $B$.  In the Schr\"{o}dinger problem, the $C$'s have as their sources the critical points on the real axis for $\lambda\neq0$.

The Richardson spectrum starts at $E=0$  with $E$ increasing through real values.  The Schr\"{o}dinger image of this must have $E_n(\lambda)$ real and therefore involve movement on the $A$'s, $B$'s and $C$'s in such a way that $E_n(\lambda)$ always increases. This Schr\"{o}dinger movement in place $(\lambda)$ is the encoding of the Richardson complication in value $(\lambda_n(E))$.

In sections \ref{sec:Rich} and \ref{sec:Diff} we have discussed the relevant differential equations in detail but for the Schr\"{o}dinger problem the potential $V(x)=-\lambda~\textrm{sgn}(x)$ is piecewise constant so that the wave function can be written down explicitly. This is the approach of Atkinson and Jabon \cite{Atkinson} which we adopt.  In the Schr\"{o}dinger problem for fixed $\lambda$ we need a function $\Delta(\lambda,E)$ that vanishes at an eigenvalue and this function is supplied by the difference of the logarithmic derivatives of the left and right eigenfunctions at $x=0$.  For $\lambda$ fixed we search for zeros in $E$ and for $E$ fixed we look for zeros in $\lambda$.  These are simple zeros but for branch points in either problem the algebraic multiplicity is 2 and we need double roots.  In Schr\"{o}dinger this requires $\Delta=\Delta_E=0$ while in Richardson we need $\Delta=\Delta_\lambda=0$.  There are numerical ways \cite{Shanley} of searching for such points while having no prior estimate of their location.  The above method has problems near points where the logarithmic derivatives are singular.  And alternative scheme is to expand the wave function in solutions of the problem when $\lambda=0$. The new $\Delta(\lambda,E)$ is then the determinant formed from the characteristic equation.

\begin{figure}[h]\center
\includegraphics{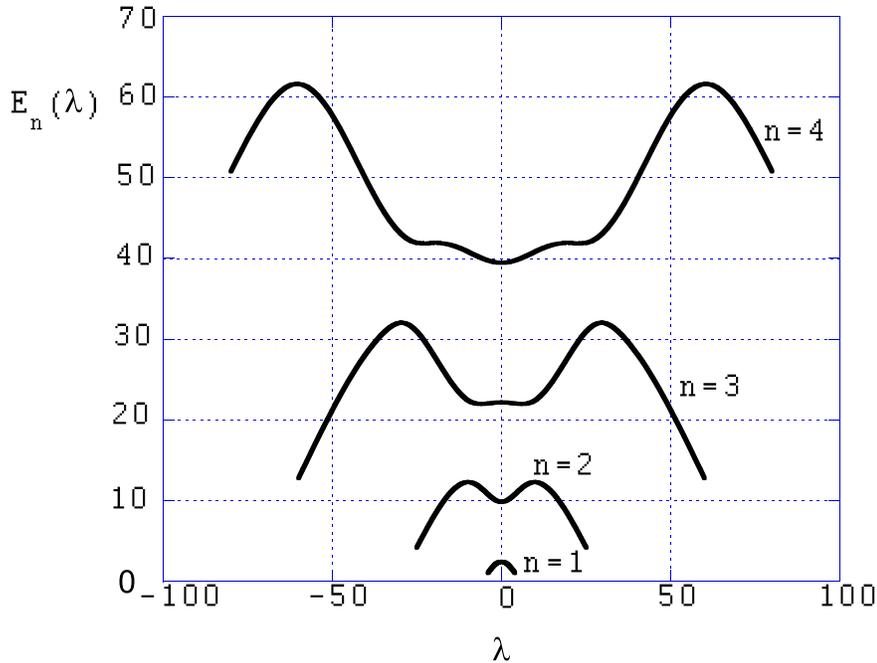}\center
\caption{\label{fig:one} $E_n(\lambda)$ versus $\lambda$ for $n=1-4$.}
\end{figure}

Given the potential $V(x)=-\lambda~\textrm{sgn}(x)$, we have solved numerically for $E_n(\lambda)$ for $n=1-4$ with results plotted in figure \ref{fig:one}.  As $|\lambda|\rightarrow\infty$, one side or the other of the well becomes infinitely deep so that $E_n(\lambda)\rightarrow-\infty$ for all $n$.  These $E_n(\lambda)$ are referred to as eigencurves by Binding and Volkmer \cite{Binding} who have shown that the $n$-th curve has $n$ local maxima. We see numerically that there are $n-1$ local minima and so it is assumed that $E_n(\lambda)$ has $2n-1$ critical points.  In \cite{Binding} it is shown that the position of the maxima in the $(\lambda,E)$ plane form a rectangular lattice so that these positions and values are known exactly. This means under spectral inversion that the information for the Richardson branch points is also known.  The minima do not form such a lattice and their positions and values have been determined numerically.

\begin{figure}[h]\center
\includegraphics{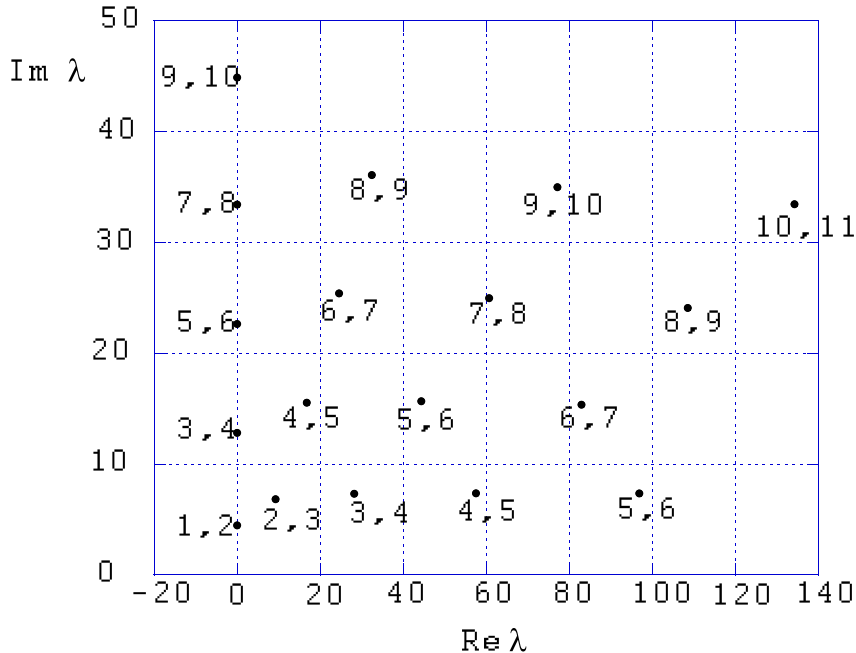}\center
\caption{\label{fig:two} Positions of branch points of $E_n(\lambda)$ for $\textrm{Re}\lambda\geq0$ and $\textrm{Im}\lambda>0$.}
\end{figure}

Binding and Volkmer \cite{Binding} have argued and to some extent proved that there are complex connections between the eigencurves of figure \ref{fig:one}.  We wish to elaborate on this and argue that all of the eigencurves are different branches of the same function and that the complex connections are facilitated by square root branch points off the real axis of $\lambda$. In order to form a Riemann surface for $E_n(\lambda)$ we make many copies of the $\lambda$ plane, label each sheet by $n$, place the $n$-th eigencurve over the real axis of the $n$-th sheet, and then search for square root branch points that couple one sheet to another.  After a lengthy numerical calculation we find the branch points shown in figure \ref{fig:two} for $|\lambda|$ and $n$ not too large. The two integers indicate the sheets that are joined at the square root.  $E_n(\lambda)$ is subject to two constraints: since $E_n(\lambda)$ is real on the real axis we have $E_n(\lambda^\ast)=E_n^\ast(\lambda)$, and in \cite{Binding} it is proved that $E_n(-\lambda)=E_n(\lambda)$. These conditions require that branch points on the imaginary axis appear as mirror pairs and also that those in the first quadrant at $\lambda=x+iy$ must appear at the four positions $\lambda=\pm x\pm iy$.  With everything in place there are $4n-2$ branch points on sheet $n$.  The singularities tend to move out with $n$ and become more numerous so it is possible that they accumulate at infinity on the asymptotic sheet.

On relevant sheets we imagine branch cuts starting at each square root and extending radially to infinity. The process of level crossing involves starting on the real axis of sheet $n$ with oscillation count $n-1$, encircling a single branch point of the type $n-n+1$, changing sheets, and then returning to the real axis of sheet $n+1$ with oscillation count $n$.  This process will play an important role in understanding the Richardson spectrum.

Some numerical values for the low-lying branch point positions and values are listed in table \ref{tab:tab1} for those on the positive imaginary axis and in the first quadrant. Using different methods, Znojil and L\'{e}vai \cite{Znojil,Znojil2} have also determined some of these values and our results are in good agreement.

\begin{table*}\center
\caption{\label{tab:tab1}Low-lying branch point for Re$\lambda\geq 0$ and $\textrm{Im}\lambda>0$.}
\begin{tabular}{ccccc}
 n-n+1&Re$\lambda$&Im$\lambda$&Re$E$&Im$E$\\ \hline
 1-2&0.0&4.475309&6.401903&0.0 \\
 2-3&9.264139&6.834853&17.617719&0.960866\\
 3-4&0.0&12.801544 &30.979714&0.0\\
 3-4&28.204239&7.318111&37.550337&1.481781\\
 4-5& 16.798312&15.527134& 52.144783&1.436416\\
 4-5&57.481587&7.358543&67.167957&1.676906
\end{tabular}\center
\end{table*}

That $E_n(-\lambda)=E_n(\lambda)$ implies that all odd derivatives vanish at $\lambda=0$ on every sheet.  This means that a real locus of type B is present on the positive and negative imaginary axes until mirror branch points are reached.  We can read off from the figure when that happens for sheets $n-n+1$. $E_n(\lambda)$ is complex above and below the branch points so that the real locus includes but then stops and these points.

\begin{figure}[h]\center
\includegraphics{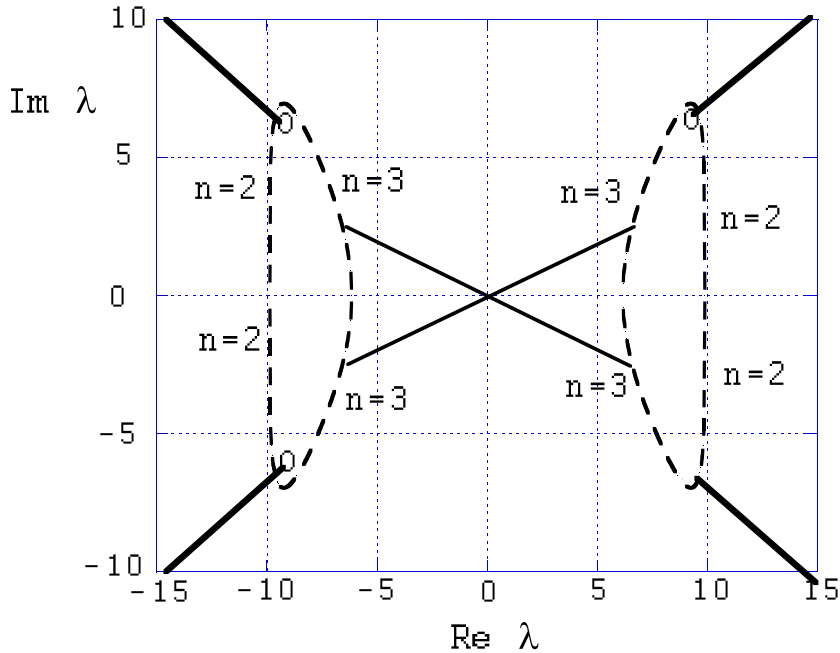}\center
\caption{\label{fig:three} Real loci of type C on sheets $n=2~\textrm{and}~3$.  The open circles are branch points from which cuts go to infinity.}
\end{figure}

The presence of critical points on the real axis for $\lambda\neq0$ implies that real loci of the type C move into the upper and lower planes.  Such points are absent on the $n=1$ sheet and we find that the next two sheets, $n=2~\textrm{and}~3$, contain a pair of real loci that are closed curves as shown in figure \ref{fig:three}.  The symmetries of $E_n(\lambda)$ imply that the curves appear as left-right mirror images. The four $n=2-3$ branch points with their radial cuts facilitate closure. These curves constitute fundamental cycles on the Riemann surface on which the eigenvalues are real.

Suppose we are on $A$ or a $B$ so that $E_n(\lambda)$ is real. The property $E_n(-\lambda)=E_n(\lambda)$ implies that there are two values of $\lambda$ with the same real energy. In Richardson this means that there are two real or two imaginary eigencouplings on different sheets for some real $E$. On a real locus of type C things are different. The end points of the two lines through the origin in figure \ref{fig:three} represent positions on the two curves with couplings $\pm\lambda$ and $\pm\lambda^\ast$.  The symmetries of $E_n(\lambda)$ imply that all four couplings have the same real eigenvalue. In Richardson we will then have four distinct complex couplings on different sheets for some real $E$. This argument works for any point on the $C$ with the exception of the real axis. We define these to be $A$-points so that only doubling occurs. To summarize, the $A$'s, $B$'s and $C$'s have reflection doubling but for the $C$'s it becomes quadrupling because of $\pm\lambda^\ast$.

\begin{figure}[h]\center
\includegraphics{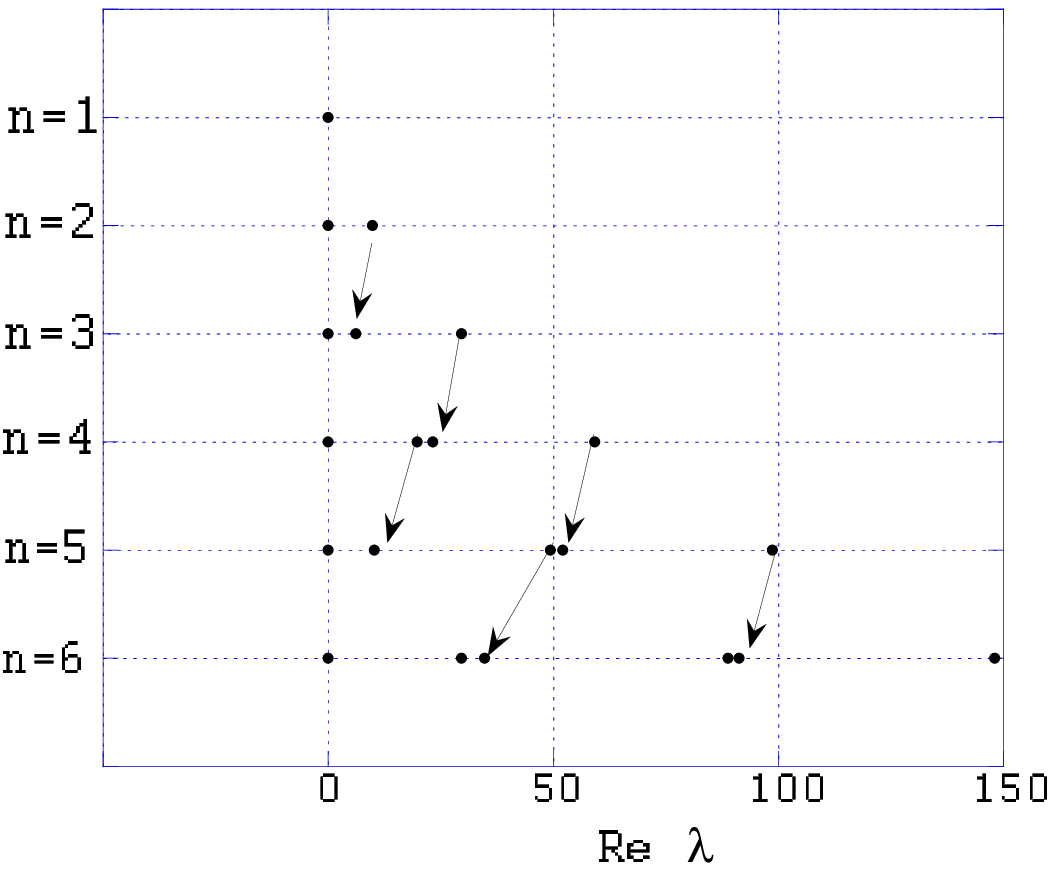}\center
\caption{\label{fig:four} Positions of critical points for $\lambda\geq0$ and $n=1-6$. The two critical points that lie on the same real locus are connected by arrows.}
\end{figure}

Since we have assumed that the number of critical points on sheet $n$ is $2n-1$, it is expected that the number of real loci will increase on the higher sheets. In figure \ref{fig:four} are shown the positions of all critical points for $\lambda\geq0$ and $n=1-6$.  The pairs that form a real locus are shown joined by an arrow. We do not display these curves but they are similar to the ones shown in figure \ref{fig:three} and branch points from the previous discussion produce the necessary level crossings that cause the curves to be closed.

There is an important difference between the locus types which is worth pointing out.  If we go to the $n=1$ sheet for some real $\lambda_0$ and look up to all sheets for that coupling, an infinite number of real eigenvalues will be present. If we repeat on the imaginary axis and if $|\lambda_0|$ is not too large, the number will also be infinite. This is somewhat indicated in figure \ref{fig:four} where the critical points at $\lambda=0$ are present vertically on all sheets. If we ask for the number for the $C$'s , a look at figure \ref{fig:four} shows that the critical points are not vertical in a similar way and a given $C$ does not appear on other sheets.  This implies that no point on a $C$ has an infinite number that is similar to the other cases.  To summarize, there are an infinite number of real eigenvalues for all $A$-points, some $B$-points, and no $C$-points.

We have argued that there are $2n-1$ critical points of $E_n(\lambda)$ for $\lambda$ real. If there were additional cases off the real axis, these would invert to become branch points of $\lambda_n(E)$ for complex $E$ which could lead to coupling crossing. Hundreds of searches using the methods described in \cite{Shanley} have been carried out which yield only critical points for real $\lambda$. This implies that Richardson branch points occur only for real values of $E$. Level crossing branch points for $\lambda$ and $E_n(\lambda)$ complex will invert to critical points in the complex domain of $E$. Since they play no important role in the Richardson problem, we have not studied them.

\section{\label{sec:Richspec}Richardson spectrum}

Here we describe the behavior of the Richardson eigencouplings as functions of the parameter $E$ and suggest a suitable Riemann surface for the problem.  Some numerical results are also presented for the eigencoupling on a few sheets.

\begin{figure}[h]\center
\includegraphics{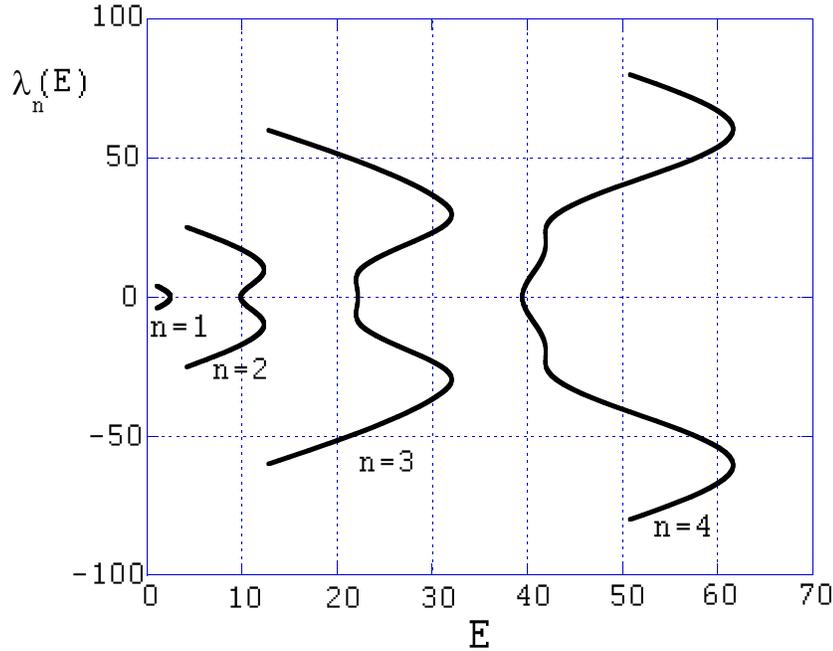}\center
\caption{\label{fig:five} $\lambda_n(E)$ versus $E$ for $n=1-4$.}
\end{figure}

Starting with the Schr\"{o}dinger eigencurves of figure \ref{fig:one} we proceed to rotate the graph and adjust the labels to get the Richardson eigencurves of figure \ref{fig:five}. In Schr\"{o}dinger each $E_n(\lambda)$ is a single-valued function of $\lambda$ but the critical points make the Richardson ones multi-valued functions of $E$. The Schr\"{o}dinger critical points with their horizontal tangents have become square roots with their vertical tangents and singular derivatives.  Since the oscillation counts are invariant under spectral inversion, these counts in effect rotate with the graph and remain unchanged.

In the Schr\"{o}dinger problem it was argued that there were complex paths that allowed passage from one eigencurve to another and we expect the same type of connections here. As $E$ becomes large and negative there will be two couplings $\pm\lambda$ for the same $E$. This double-valuedness can be resolved by putting each branch on a different sheet so that we extend our notation, $\lambda_n(E)\rightarrow\lambda_n^\pm(E)$, where $n$ is assigned by virtue of the oscillation count $(n-1)$ when $E<0$ and $\pm$ indicates the sign of the coupling  in that region. As $E$ becomes positive, the inverted Schr\"{o}dinger critical points become square roots that couple one sheet to another.  In figure \ref{fig:five} it is seen that the functions $\lambda_n^+(E)$ and $\lambda_n^-(E)$ have a common branch point at $E=n^2(\pi^2/4)$ where the two couplings vanish.  These are the images of the Schr\"{o}dinger problem at $\lambda=0$.  As in the Schr\"{o}dinger case we expect an infinite number of sheets.  By counting sheets or singularities it seems likely that each of the Riemann surfaces has infinite genus, and speaking crudely, spectral inversion takes one of them and turns it inside out to form the other.

To guide our results when the eigencouplings leave the real domain, we will use a theorem (5.8.2) of Zettl \cite{Zettl} that gives a bound on the number of complex eigencouplings.  That the Richardson problem is right-indefinite is not sufficient to guarantee complex results.  It is necessary that the operator implicit on the left of the equation (\ref{eq:Richa}) be negative-definite; such a situation is termed left-indefinite and there is a simple test for it.  Call Richardson with weight $w(x)=-\textrm{sgn}(x)$ the $w(x)$ problem and, if a related problem with weight $|w(x)|$ has $m$ negative eigencouplings, then the $w(x)$ problem can have at most $2m$ non-real eigencouplings.  The $|w(x)|$ problem is right-definite and reads

\vspace{0.3cm}
\begin{eqnarray}
-\phi_n^{''}(x)+[\lambda_n(E)+E\,]\phi_n(x)=0\\
~\phi_n(\pm1)=0,\nonumber\\ \nonumber
\label{eq:Rich1}
\end{eqnarray}
where $\lambda_n(E)$ is the eigencoupling of the $|w(x)|$ problem and the solution is $\lambda_n(E)=n^2(\pi^2/4)-E$.  The first value of $E$ that allows complex results is one just to the right of $\pi^2/4$ which is as expected and is where the $1^+$ and $1^-$ curves meet at their branch point.  Note that the theorem confirms that nothing interesting happens for $E<\pi^2/4$.  In the general case if $E$ is just to the right of $n^2(\pi^2/4)$, the bound is $2n$.

\begin{figure}[h]\center
\includegraphics{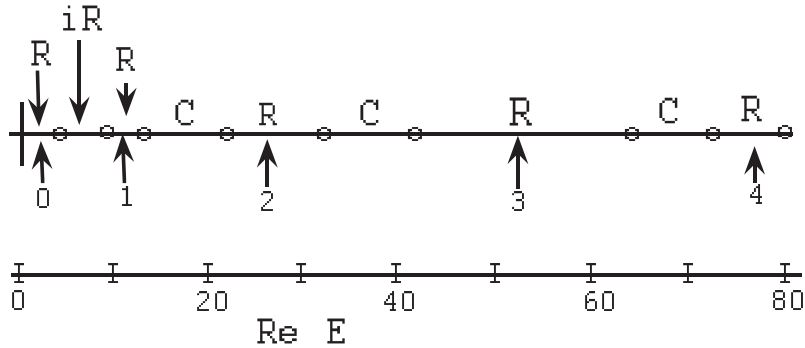}\center
\caption{\label{fig:six} $\lambda_1^+(E)$ versus $E$.  The symbols denote real (\texttt{R}), imaginary(\texttt{iR}), and complex (\texttt{C}). The integers give the oscillation count.}
\end{figure}

We now show results in figure \ref{fig:six} for $\lambda_1^+(E)$ which was originally studied in \cite{Atkinson}.  Rather than trying to plot complex functions, we  only show whether the result is real ($\mathbb{R}$), imaginary ($i\mathbb{R}$), or complex ($\mathbb{C}$).  It is seen that there are short segments in $E$ in which the three spectral types occur over and over and this continues past that shown in the graph.  The boundaries between segments are square root singularities that arise from inversion of Schr\"{o}dinger critical points.  The integer below each real segment is the oscillation count in that interval.  That this changes from one section to another was considered a puzzle in earlier work \cite{Atkinson}; whenever the eigencoupling returns from its complex plane, the eigenfunction comes back with another node.  There are two ways of clarifying the puzzle.   The Richardson way is to recall that the eigencurves of figure \ref{fig:five} inherited the oscillation counts from the Schr\"{o}dinger curves. The behavior of $\lambda_1^+(E)$ represents real results that would be visible on eigencurves separated by non-real segments that apparently induce a change from one eigencurve to the next one on the right. If viewed only in the real domain, this would appear as jumping from one curve to the next, with a unit increase in oscillation count.  In the Schr\"{o}dinger view there is a more natural mechanism for this curve jumping and that is the process of level crossing.  The Schr\"{o}dinger image of the Richardson segments would appear on figure \ref{fig:one} and involve movement on the $A$'s, $B$'s and $C$'s so that level crossing must occur between the real segments and this causes the needed change in oscillation count.  The Richardson jumping occurs horizontally in figure \ref{fig:five} with $E$ real and increasing, while Schr\"{o}dinger jumping occurs vertically in figure \ref{fig:one} with $E_n(\lambda)$ real and increasing.  In either case a jump of curves involves a unit increase in oscillation count.

The eigencoupling $\lambda_1^-(E)$ need not be shown on a separate graph since the plot would not differ from that seen in figure \ref{fig:six}.  In the real region of $E<\pi^2/4$, we have by construction $\lambda_1^-(E)=-\lambda_1^+(E)$ and we find this to be true also in the complex domain and also for other $n$ that have been studied.  This is the realization of the idea discussed in section \ref{sec:Schro} that for $\lambda$ on $A$, $B$, or $C$, there will be two couplings of opposite sign that occur on two different Richardson sheets.

\begin{figure}[h]\center
\includegraphics{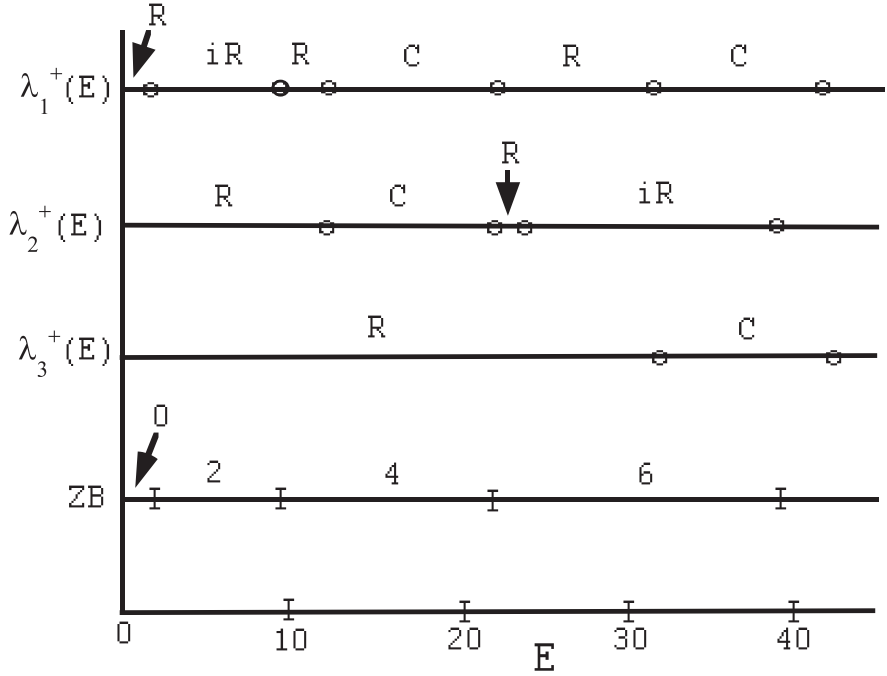}\center
\caption{\label{fig:seven} The spectral types for $\lambda_n^+(E)$ for $n=1-3$ versus $E$.  ZB is the Zettl bound.}
\end{figure}

We combine in figure \ref{fig:seven} results for $\lambda_1^+(E)$, $\lambda_2^+(E)$, and $\lambda_3^+(E)$ together with the Zettl bound (ZB) on the number of complex results.  To the left of $\pi^2/4$, the bound is zero and all results are real.  In addition to $\lambda_n^+(E)$ for $n=1-3$, there is a $\lambda_n^-(E)$ that is not shown and has opposite sign. In the first imaginary $(i\mathbb{R})$ segment the two couplings $\lambda_1^\pm(E)$ saturate the bound at 2 and all other results are real. At $E$ just above $\pi^2$, everything turns real and the bound is not achieved for a short interval.  At $E$ just above $5\pi^2/4$, the $1^\pm$ and $2^\pm$ segments become complex with the $1^\pm$ contributing $\pm\lambda$ and the $2^\pm$ giving $\pm\lambda^\ast$ and the bound of 4 is reached.  Since the Schr\"{o}dinger image of these intervals all involve $n=2~\textrm{and}~3$, they are the Richardson version of the points singled out on the real loci of type $C$ of figure \ref{fig:three} where we anticipated four couplings for some real $E$. This scheme continues on as $E$ increases and more couplings become complex.

It is useful to compare the complexity of the Richardson spectrum with what would be observed in the Schr\"{o}dinger problem.  There as real $\lambda$ varies we will have one spectral type ($\mathbb{R}$), no change in oscillation count, and no square roots. The Schr\"{o}dinger problem appears to be much simpler but our overall task was to show that the two functions are inverses.  This seeming paradox is avoided by observing that the quoted Schr\"{o}dinger simplicity is all in the domain of the real loci of type $A$.  It will turn out that Schr\"{o}dinger is sufficiently complicated on the $B$'s and $C$'s so that no conflict exists.

\section{\label{sec:Spect}Spectral inversion of Richardson}

We have argued above that the Schr\"{o}dinger $A$'s, $B$'s, and $C$'s with their real values and complicated places become the Richardson spectrum after inversion with its real places and complicated values.  This argument does not specify which real locus is the image of a certain Richardson segment.  The details of this will be worked out below in a few cases.

Suppose we could go to the Schr\"{o}dinger problem and erase the $B$'s and $C$'s from existence so that only the $A$'s remain.  The modified Richardson problem would then have only real eigencouplings and the system would lose much of its interest.  We conclude that the following two statements are equivalent: the Schr\"{o}dinger problem has $B$'s and $C$'s and the spectrum of Richardson is complicated.  This is consistent with our general argument that the two types of complication are the inversions of each other.

\begin{figure}[h]\center
\includegraphics{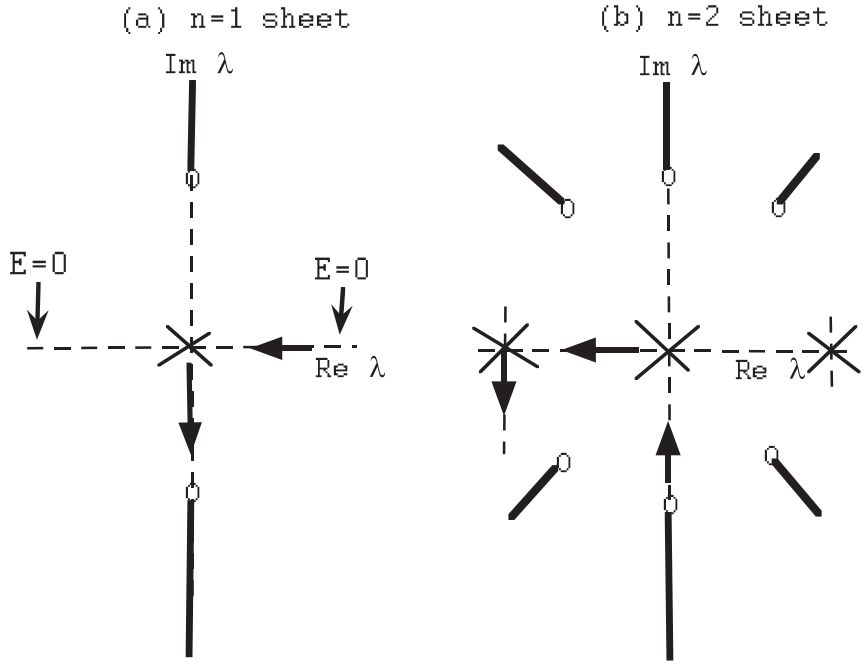}\center
\caption{\label{fig:eight} Riemann surface of $E_n(\lambda)$. (a) $n=1$ sheet, (b) $n=2$ sheet.  The broad lines are branch cuts.}
\end{figure}

The first four segments of the Richardson spectrum in figure \ref{fig:six} are $\mathbb{R}$, $i\mathbb{R}$, $\mathbb{R}$, $\mathbb{C}$ with the oscillation count being 0 and then 1 in the real segments and with the segment boundaries being square roots.  In figure \ref{fig:eight} are shown the $n=1$ and $n=2$ Riemann sheets on which $E_n(\lambda)$ takes its values. These sheets have a pair of common branch points on the imaginary axis while sheet $n=2$ has four branch points leading to sheet $n=3$.  Every double-x locates a critical point.  To match the Richardson sequence of values we need a Schr\"{o}dinger sequence of places $\lambda\in A,~B,~A,~C,$ with the boundaries being critical points, the oscillation count changing from 0 to 1, and with $E_n(\lambda)$ always increasing. The point labeled $E=0$ on the positive real axis of the $n=1$ sheet is the image of $\lambda_1^+(E)$ at $E=0$. If we follow the arrows on the two sheets, we find the sequence of places $\lambda\in A,~B,~A,~C,$ the boundaries are critical points, and $E_n(\lambda)$ is always increasing.  When the branch point is reached on Im$\lambda<0$ on the $n=1$ sheet, the first instance of level crossing occurs where we do not encircle a branch point, we simply touch it, change sheets, $E_n(\lambda)$ continues to increase and when the origin is reached on sheet $n=2$, the oscillation count will be unity as required.  The origin is a minimum so that we can go left until a maximum is reached.  This maximum is a critical point so that a real locus of type $C$ is available and the image point moves onto it and $E_2(\lambda)$ remains real and increasing. This completes the required sequence $\lambda\in A,~B,~A,~C$.

\begin{figure}[h]\center
\includegraphics{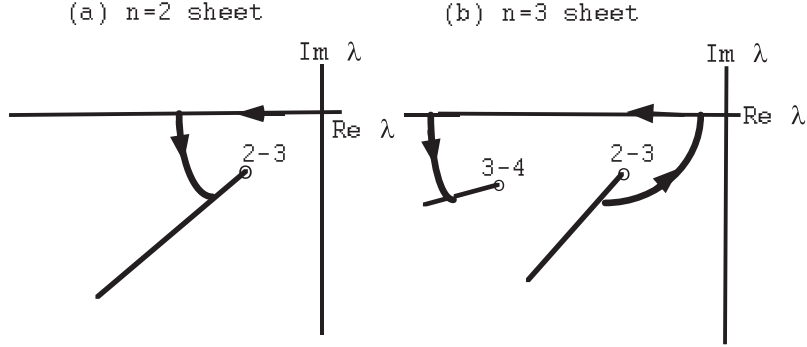}\center
\caption{\label{fig:nine} Schematic Schr\"{o}dinger image of the Richardson function $\lambda_1^+(E)$. (a) $n=2$ sheet, (b) $n=3$ sheet.}
\end{figure}

Going back to the origin of the $n=2$ sheet, the Schr\"{o}dinger sequence  $\lambda\in A,~C,~A,~C,\cdots$ must be achieved to match the Richardson result of figure \ref{fig:six}.  The way that this is accomplished is illustrated schematically in figure \ref{fig:nine} (a) for the $n=2$ sheet and in figure \ref{fig:nine} (b) for $n=3$. The origin for $n=2$ is a minimum, we proceed left to the maximum, move onto the real locus of type $C$, the path encloses a branch point of type $n=2-3$, we pass to sheet $n=3$, and when we return to the real axis in figure \ref{fig:nine} (b), the oscillation count will be 2 and we will again be at a minimum.  This sequence of steps is repeated over and over leading to larger $n$ and oscillation count.  This is our qualitative argument that the two functions are inverses of each other. Other $\lambda_n^+(E)$ could be studied in a similar way.

The Schr\"{o}dinger sequence of the places $\lambda\in A,~B,~A,~C,~A,~C,\cdots$ is the realization of the curve jumping discussed in section \ref{sec:Richspec}.  When viewed only in the real domain in figure \ref{fig:one}, we would see the sequence $A(n=1),~A(n=2),~A(n=3)$ with the invisible level crossing sequence $B$, $C$, $C$ providing the mechanism for getting from one eigencurve to the next.  In the Richardson problem there is nothing as specific as level crossing  to account for the change in oscillation count so that we rely on the Schr\"{o}dinger description and argue that the count is invariant under spectral inversion.

Here we add few additional comments.  A real Richardson segment may not contain a critical point since that would imply that the Schr\"{o}dinger problem is singular for real $\lambda$. On an imaginary Richardson segment there is no such restriction and $\lambda_1^+(E)$ must have a critical point that is the image of the branch point in figure \ref{fig:eight} (a) for Im$\lambda<0$.  This critical point differs from the Schr\"{o}dinger examples since two imaginary loci on which Re$\lambda_1^+(E)=0$, intersect orthogonally.  In this way the local information on the two Schr\"{o}dinger sheets near the square root gets unwound and inverted and appears on one Richardson sheet. On the neighboring sheet containing $\lambda_1^-(E)=-\lambda_1^+(E)$, there is a similar critical point at the same value of E that is the image of the branch point for Im$\lambda>0$.  The Schr\"{o}dinger image of the function $\lambda_1^-(E)$ starts on the left at $E=0$ on the $n=1$ sheet and always moves in directions opposite to those of the arrows.

It is amusing to conjecture that $\lambda_1^+(E)$ continues its oscillation $\mathbb{R}$, $\mathbb{C}$, $\mathbb{R}$, $\mathbb{C}, \cdot$ all the way to $E=\infty$.  The Schr\"{o}dinger image of this would be an oscillation $A$, $C$, $A$, $C,~\cdot$ all the way to $n=\infty$ and Re$\lambda=-\infty$.  In this way the information on the real axis of one Richardson sheet would appear on an infinite number of Schr\"{o}dinger sheets. If we single out the point where $E_1(\lambda)=0$ on the right side of the $n=1$ sheet and also the point $\lambda=-\infty$ on the asymptotic sheet, then these points would be connected by a very long real locus that is the image of positive Richardson real line.

\section{\label{sec:Oscil}Oscillation counts and defective multiplicity}

\noindent This section focuses on the observation - quite familiar by now - that the Richardson problem is more complicated for real $E$ than the Schr\"{o}dinger one is for real $\lambda$. To examine this further, two questions are posed.

\begin{enumerate}\item For given real $E$, how many Richardson eigenfunctions have oscillation count $m$?\\
                 \item How many coincident square root singularities of the eigencoupling are there over
                 a given real value of $E$?\\
\end{enumerate}
If we ask these questions of Schr\"{o}dinger quantities for a given real $\lambda$, the very simple answers are: one for all $m$, and zero.  Figures \ref{fig:one} and \ref{fig:five} contrast the singled-valued simplicity of Schr\"{o}dinger with the multi-valued complexity of Richardson.

We have no general result for the first question but we will work through some particular cases.  For some real $E$ let $N_m$ be the number of Richardson eigenfunctions with oscillation count $m$.  In figure \ref{fig:five} if we approach a vertical tangent from the side with two real branches, there will be two eigenfunctions with the same count which limits to a single one at the branch point since the geometric multiplicity is unity.  Some special cases are given in the following list that can be followed in figure \ref{fig:five}.

\begin{tabbing}
\hspace*{1cm}$E<\pi^2/4\approx2.47$ \=\hspace*{2cm} $N_m=2,m\geq0$\\
\hspace*{1cm}$E=\pi^2/4$ \>\hspace*{2cm} $N_0=1;N_m=2, m\geq1$\\
\hspace*{1cm}$\pi^2/4<E<\pi^2\approx9.87$ \>\hspace*{2cm} $N_0=0;N_m=2, m\geq1$\\
\hspace*{1cm}$E=\pi^2$ \>\hspace*{2cm} $N_0=0;N_1=3;N_m=2, m\geq2$\\
\hspace*{1cm}$\pi^2<E<5\pi^2/4\approx12.34$ \>\hspace*{2cm} $N_0=0;N_1=4;N_m=2, m\geq2$\\
\hspace*{1cm}$E=5\pi^2/4$ \>\hspace*{2cm} $N_0=0;N_1=2;N_m=2, m\geq2$\\
\end{tabbing}
We stop here because further cases involve numerical estimates for minima. Notice that $N_1$ goes through the sequence $2,3,4,2$ and $N_1=0$ after this. Things get very complicated as $E$ increases and obtaining a general result seems difficult.

In the Schr\"{o}dinger problem it is not a significant event  if there are two distinct couplings that are critical points with the same energy.  Many cases can be seen in figure \ref{fig:one} and they arise from $E_n(-\lambda)=E_n(\lambda)$ at critical points. Such an event when viewed in the Richardson problem is more significant mathematically since the eigencoupling will have coincident branch points on different sheets.  The task posed by the second question is to count this number for given $E$. In dealing with this it is more convenient to use the Schr\"{o}dinger description at critical points rather than the Richardson one at square roots. In problems of this sort these critical point positions and values are usually undecipherable numbers that are presumed irrational and can only be estimated numerically.  The existence of the Binding and Volkmer \cite{Binding} lattice of maxima means that we can reduce this part of the problem to integers, after dividing out a common irrationality. Apart from those at $\lambda=0$, we have no model for the minima and their positions and values are found to be in the undecipherable class. It follows that only about half of the problem can be treated exactly. Some critical point positions and values for minima with $\lambda\neq0$ are listed in table \ref{tab:tab2}.

\begin{table*}[h]\center
\caption{\label{tab:tab2}Position of minima of $E_n(\lambda)$ for $\lambda\neq0$.}
\begin{tabular}{ccc}
 n&$\lambda$&$E_n(\lambda)$\\ \hline
 3&$\pm$6.151546&21.996039 \\
 4&$\pm$23.271272&41.909383 \\
 5&$\pm$10.258305&61.478860 \\
 5&$\pm$52.084097&71.536568 \\
 6&$\pm$34.746014&91.264186 \\
 6&$\pm$91.259508&111.019851
\end{tabular}\center
\end{table*}

In dealing with the second question we call the number of coincident square roots the defective multiplicity $N_d$. Note that over almost all $E$, $N_d=0$ because no square roots are present.  The first unexpected case that might be called ``accidental" occurs when the energy at the Schr\"{o}dinger outer maxima for $n=4$ at $\lambda=\pm6\pi^2$ coincides with that for the central maximum of $E_5(0)=25(\pi^2/4)$, leading to $N_d=3$ in Richardson.  Because of symmetry, non-vanishing $N_d$ will be even and at least 2 unless a value at $\lambda=0$ is involved, as it is here.

\begin{figure}[h]\center
\includegraphics{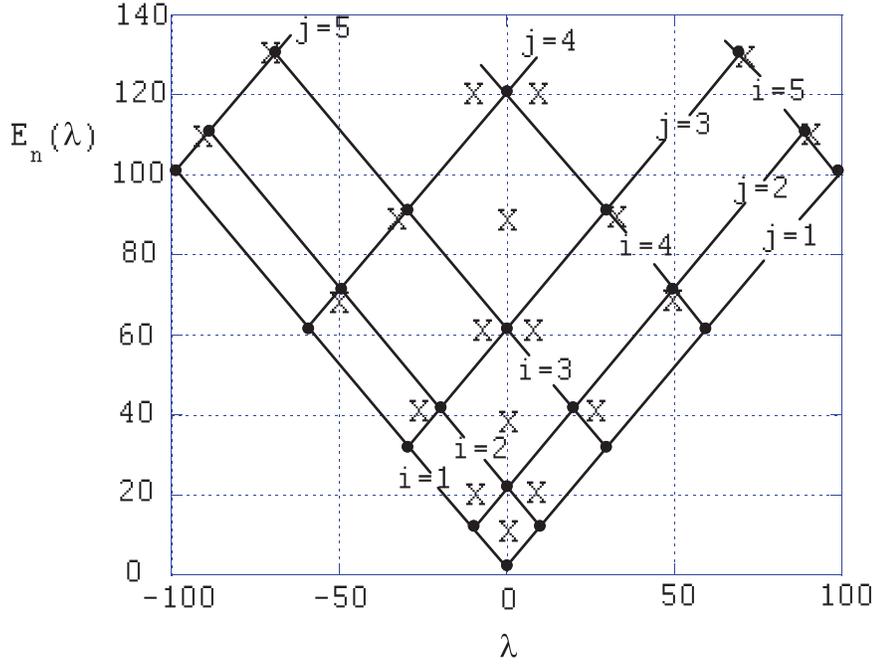}\center
\caption{\label{fig:ten} Positions of extrema of $E_n(\lambda)$ versus $\lambda$. The solid dots are maxima and the crosses are minima. The lattice coordinates are indexed as $(i,j)$.}
\end{figure}

To study further examples we set up the lattice of maxima shown as dots in figure \ref{fig:ten} with the minima shown as crosses. Lattice points are located by the pair of positive integers $(i,j)$ giving the couplings $\lambda_{ij}$, the energies $E_{ij}$ and the quantum numbers $n_{ij}$ as follows

\parbox{9cm}{\begin{eqnarray*}
\lambda_{ij}=(\pi^2/4)[2(i+j-1)(j-i)],\\
E_{ij}=(\pi^2/4)[2i^2+2j^2-2i-2j+1),\\
n_{ij}=i+j-1.\end{eqnarray*}}\hfill\parbox{3.5cm}{\begin{eqnarray}\end{eqnarray}}\vspace{0.3cm}
Dividing out the irrationality gives integer values for the reduced couplings and energies

\begin{eqnarray}
\tilde{\lambda}_{ij}&=&2(i+j-1)(j-i)=\textrm{even integer},\\
\tilde{E}_{ij}&=&2i^2+2j^2-2i-2j+1=\textrm{odd integer}.\\ \nonumber
\label{eq:Oscill1}
\end{eqnarray}
The defective multiplicity is the number of distinct pairs of positive integers $(i,j)$ that have the same $\tilde{E}_{ij}$. We can form a symmetric matrix $\tilde{E}$ whose first few elements are

\begin{eqnarray}
\tilde{E}=\left[\begin{array}{llll}
1&5&13&25\\
5&9&17&29\\
13&17&25&37\\
25&29&37&49\end{array}\right]\\ \nonumber
\end{eqnarray}
Note that $N_d=0$ for any real number that is not present in $\tilde{E}$, at least as far as maxima are concerned. By inspection, $N_d=1$ for $\tilde{E}_{ij}=1,9,49$; $N_d=2$ for $\tilde{E}_{ij}=5,13,17,29,37$; and $N_d=3$ for $\tilde{E}_{ij}=25$, which is the case we called accidental. In general we must have $N_d\geq2$ for $i\neq j$ and $N_d\geq1$ and odd if $i=j$.

A classical problem in number theory asks how many positive, negative or vanishing integer pairs $(i,j)$ lead to the same $N=i^2+j^2$.  The well-known solution \cite{Moreno} involves determining all the integer factors that divide $N$ with no remainder. The even divisors are ignored and the odd ones are separated into two classes

\parbox{9cm}{\begin{eqnarray*}
n_1(N)=1~\textrm{mod}~4=1,5,9,13,\ldots,\\
n_3(N)=3~\textrm{mod~}4=3,7,11,15\ldots.\\\end{eqnarray*}}
\hfill\parbox{3.5cm}{\begin{eqnarray}\end{eqnarray}}
The classical result for the number of pairs is $4[n_1(N)-n_3(N)]$. Our problem differs because only positive integers are involved and our expression for $\tilde{E}_{ij}$ is a circle centered at $(i,j)=(1/2,1/2)$ rather than one centered at the origin. By studying hundreds of examples we find empirically that a similar result is obtained

\vspace{0.3cm}
\begin{equation}
N_d=n_1(\tilde{E}_{ij})-n_3(\tilde{E}_{ij}).
\end{equation}\vspace{0.3cm}
For example, the factors of $\tilde{E}_{ij}=25$ are $1,5~ \textrm{and} ~25$ which are all congruent to $1~\textrm{mod}~4$ so that $N_d=3$, as has already been seen. If we return to the matrix $\tilde{E}$ and choose an odd integer that is not present such as $15$, we obtain $n_1(15)=n_3(15)=2$ so that $N_d=0$ is verified since the energy is not on the lattice.

As we consider larger $\tilde{E}_{ij}$, the first appearance of a certain $N_d$ is found to grow slowly. Table \ref{tab:tab3} gives the smallest value of $\tilde{E}_{ij}$ that produces the listed values of $N_d$ and the first three entries have already been discussed. It appears that $N_d$ goes slowly to infinity with $\tilde{E}_{ij}$. The last entry with $N_d=10$ involves maxima with coordinates $(i,j)=(6,64), (13,63),(28,58),(30,37),(43,48)$, and the number doubles by exchanging factors. The quantum numbers involved are $n=69,75,85,86~\textrm{and}~90$. The total number of maxima for the eigencurves is 405 and of these, 10 are coincident at $\tilde{E}_{ij}=8125$. These curves must oscillate in a very complicated manner so that, although they enter each others vicinity, they must do so without intersection.

To this point we have only studied $N_d$ that arise from coincident Schr\"{o}dinger maxima of the form (max,max) but we have ignored the types (max, min) or (min, min). The only exact information we have about minima is that at $\lambda=0$, the reduced energies are even integers with $\tilde{E}_n(0)=n^2$, with $n=2,4,6\cdots$. Since at maxima $\tilde{E}_{ij}=\textrm{odd integer,}$ (min, max) degeneracy of this type is not possible. For $\lambda\neq0$, a (min,max) degeneracy requires $\tilde{E}(\lambda)=$ odd integer at the minimum. For $n\leq10$, we find that this does not happen and results at the minima are all of the undecipherable type. There will be many degeneracies of the (min,min) type due to $E_n(-\lambda)=E_n(\lambda)$. To find others would involve problems with numerical round-off error. We are confident that there are none of this type for $n\leq10$.

\begin{table*}\center
\caption{\label{tab:tab3}Smallest $\tilde{E}_{ij}$ that has defective multiplicity $N_d$.}
\begin{tabular}{ccccc}
$\tilde{E}_{ij}$&$N_d$&&$\tilde{E}_{ij}$&$N_d$\\ \cline{1-2}\cline{4-5}
 1&1&&325&6 \\
 5&2&&15625&7 \\
 25&3&&1105&8 \\
 65&4&&4225&9 \\
 625&5&&8125&10
\end{tabular}\center
\end{table*}

\section{\label{sec:Sum}Summary}
\noindent In this paper we have studied the eigenvalues of two SL problems, each dependent  on a parameter, and shown that they are inverse functions of each other. This has allowed an understanding of the mysterious features of the Richardson spectrum in terms of Schr\"{o}dinger behavior in the complex domain of the coupling parameter. In Schr\"{o}dinger the properties of $E_n(\lambda)$ on the three types of real locus were shown to be related to the three spectral types of Richardson. The mathematical features that generate the complexities in each description are the critical points and square root branch points of the eigenquantites. In Schr\"{o}dinger the critical points act as sources of the $B$'s and $C$'s while the branch points tie the Riemann sheets together and facilitate the level crossing that accounts for the oscillation count variability of the Richardson spectrum. The spectral inversion  of the Schr\"{o}dinger critical points yields Richardson square roots that separate the three spectral types and tie those Riemann sheets together

The complications in each problem are driven by the simple feature that the Richardson weight, or equivalently the Schr\"{o}dinger potential, changes sign once. We saw in section \ref{sec:Diff} that these complexities were absent for problems with no sign change

If asked to explain the Richardson spectrum on its own terms and with no reference to another problem, we do not know how to proceed. It may be best to think of the two problems as a closely related pair, with Schr\"{o}dinger contributing to the understanding of Richardson.

\section*{References}
\bibliography{Shanley}

\end{document}